\documentclass[a4paper,onecolumn,11pt,accepted=2017-05-09]{quantumarticle}
\pdfoutput=1
\usepackage[utf8]{inputenc}
\usepackage[english]{babel}
\usepackage[T1]{fontenc}
\usepackage{amsmath}
\usepackage{hyperref}
\usepackage{subcaption}
\usepackage{amssymb}
\usepackage{tikz}
\usepackage{lipsum}
\definecolor{BrickRed}{RGB}{203, 65, 84}
\definecolor{mycolor}{HTML}{90E0EF}
\hypersetup{
	colorlinks=true,
	citecolor= BrickRed,
	linkcolor=BrickRed,
	filecolor=BrickRed,      
	urlcolor= BrickRed,
	pdftitle={Overleaf Example},
	pdfpagemode=FullScreen,
}
\begin{document}

\title{Quantum-walk search in motion}

\author{Himanshu Sahu}
\affiliation{Department of Instrumentation \& Applied Physics, Indian Institute of Sciences, C.V. Raman Avenue, Bangalore, 560012, India}
\orcid{0000-0002-2445-2701}
\email{himanshusah1@iisc.ac.in}

\author{Kallol Sen}
\email{kallolmax@gmail.com}
\orcid{0000-0002-5540-0285}
\affiliation{ICTP South American Institute for Fundamental Research, IFT-UNESP ($1^{\circ}$ andar), Rua Dr. Bento Teobaldo Ferraz 271, Bloco 2 - Barra Funda 01140-070 São Paulo, SP Brazil}
\affiliation{Quantum Information and Computing (QuIC) laboratory, Light and Matter Physics, Raman Research Institute, C. V. Raman Avenue Sadashivanagar, Bangalore, 560012, Karnataka, India}

\maketitle

\begin{abstract}
    In quantum computing, the quantum walk search algorithm is designed for locating fixed marked nodes within a graph. However, when multiple marked nodes exist, the conventional search algorithm lacks the capacity to simultaneously amplify the marked nodes as well as identify the correct chronological ordering between the marked nodes, if any. To address this limitation, we explore a potential extension of the algorithm by introducing additional quantum states to label the marked nodes. The labels resolve the ambiguity of simultaneous amplification of the marked nodes. Additionally, by associating the label states with a chronological ordering, we can extend the algorithm to track a moving particle on a two-dimensional surface. Our algorithm efficiently searches for the trajectory of the particle and is supported by a proposed quantum circuit. This concept holds promise for a range of applications, from real-time object tracking to network management and routing. 
\end{abstract}

\section{Introduction}

Quantum computers are engineered with the purpose of surpassing the computational capabilities of conventional computers through the execution of quantum algorithms\,\cite{bacon_recent_2010,montanaro_quantum_2016,ladd_quantum_2010}. These quantum algorithms have a wide range of practical applications from cryptography, search and optimization, and quantum system simulation to the resolution of extensive sets of linear equations \,\cite{gisin_quantum_2002,magniez_search_2007,cerezo_variational_2021,shor_algorithms_1994,lloyd_universal_1996,harrow_quantum_2009,childs_quantum_2010,doi:10.1126/science.1217069,bauer_quantum_2020,ambainis_quantum_2004,georgescu_quantum_2014,trabesinger_quantum_2012}. Notably, Grover's search algorithm stands out as a widely recognized quantum algorithm capable of searching unsorted databases with a quadratic speed advantage over its classical counterparts \,\cite{grover_fast_1996}. Grover's algorithm, combined with quantum walk, has motivated the foundation and development of the commonly known research venture of ``Quantum Walk Search Algorithms'' for searching and sorting unstructured spatial data\,\cite{santos_szegedys_2016,magniez_search_2007,santha_quantum_2008,portugal_quantum_2013}. 

Unlike classical random walks, quantum walks\, \cite{venegas-andraca_quantum_2012,ambainis_quantum_2003} represent the walker by a quantum state, thus allowing a probabilistic interpretation of the particle's dynamics. Quantum walks are a successful framework for modeling controlled dynamics in quantum systems\,\cite{oka_breakdown_2005,engel_evidence_2007,mohseni_environment-assisted_2008,PhysRevA.78.022314,chandrashekar_disordered-quantum-walk-induced_2011,kitagawa_exploring_2010}, building quantum algorithms and shown to constitute a universal model of quantum computation\,\cite{ambainis_quantum_2003,shenvi_quantum_2003,Childs_2003,ambainis2004coins}. Depending on the premise and specific applications being addressed, there are two classes of quantum walks {\it viz.} discrete-time quantum walks (DTQWs) and continuous-time quantum walks (CTQWs) that are broadly employed.

The Hilbert space of DTQWs comprises two components: coin space and position space. Coin space represents the internal state of the walker, while position space defines the lattice structure in which the walker moves. The evolution of a particle in a DQTWs is governed by a sequence of unitary operators, each composed of a coin operator and a shift operator. The coin operator manipulates the internal state of the walker, while the shift operator displaces the walker to an adjacent vertex on the lattice, conditioned by the walker's internal state\,\cite{chandrashekar_discrete-time_2010,venegas-andraca_quantum_2012}. In contrast, CTQWs describes the state evolution as a continuous function of time by associating a Hamiltonian to the unitary operator $U=\exp(-i H t)$ \,\cite{venegas-andraca_quantum_2012}.

Quantum walk-based search algorithms  (QWSA) have diverse applications across fields such as optimization, machine learning, cryptography, and network analysis\,\cite{kadian_quantum_2021}. Quantum search algorithms have an asymptotic quadratic acceleration in terms of oracle calls unlike their classical cousins\, \cite{ambainis_quantum_2003, santos_szegedys_2016}. \\

\begin{figure}
	\begin{minipage}[c]{0.65\textwidth}
		\includegraphics[width=\textwidth]{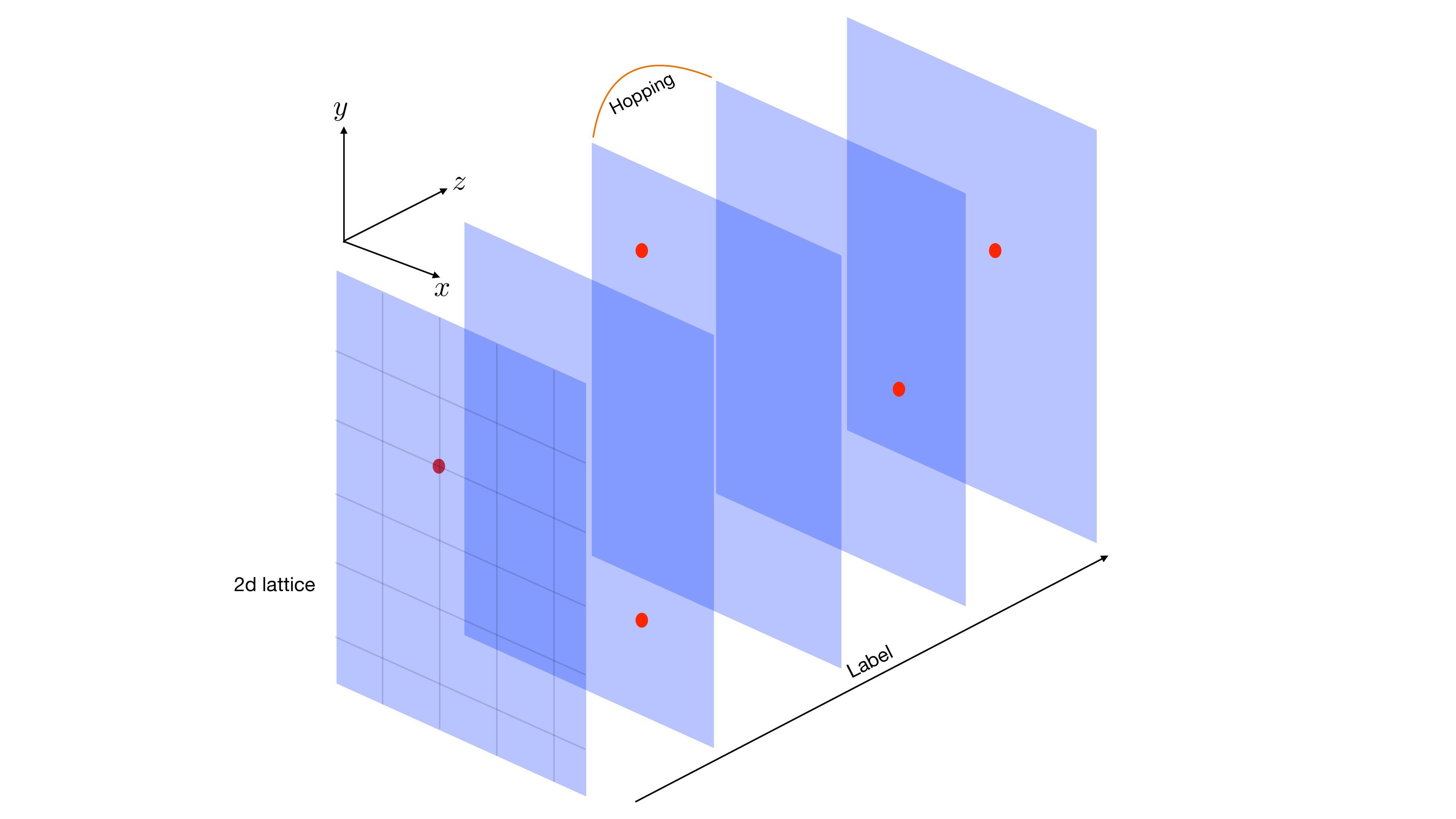}
	\end{minipage}\hfill
	\begin{minipage}[c]{0.35\textwidth}
	\caption{Replication of the $2d$ lattice ($xy$- grid) on the $z$-direction which denotes the labels. The original search problem, including five marked nodes on a $2d$ lattice, reduces to the search problem, including five $2d$ sheets with one marked node each. We can allow for hopping between the sheets to enhance the probability distribution for the marked nodes at the cost of diminishing the probability for the unmarked ones.}
	\label{fig:labelled_node}
	\end{minipage}
\end{figure}

\noindent Quantum-walk search with multiple points has been extensively studied in previous literature\,\cite{xu_robust_2022,wong_exceptional_2017,j_spatial_2018,li_generalized_2020,glos_upperbounds_2021,PhysRevA.103.062202}. While the algorithm can locate multiple nodes within a graph, however not all the marked points are equally amplified. Further, a chronological ordering of the marked nodes, if there exists any, is completely ignored by the algorithm. These drawbacks make the algorithm unsuitable for sorting dynamical data. In the present work, we specifically address these issues regarding the QWSA. To resolve this, we associate a label to each of the marked nodes in the form of additional quantum bits. While this increases the dimension of the associated Hilbert space, we secure an extra handle that can be used to distinguish the ordering of marked nodes. We study the specific case of QSWA in finite two-dimension lattice with open and periodic boundary conditions {\it, i.e.} the open grid and the torus. In addition, we also consider the instances where the labels can be static or dynamic, as explained diagrammatically in Fig.~\ref{fig:labelled_node}.


The distinction between static and dynamic labeling relies on whether there is a flow of probability between various sheets. Our refined algorithm can address a variety of applications, including real-time object tracking, trajectory prediction, financial market analysis, dynamic optimization problems, and network management and routing that includes dynamical components. The labeling concept is not new and has been applied to element distinctness problems in quantum algorithms. However, to our knowledge, we are not aware of the application of the labeling concept in the context of quantum search algorithms. As a concrete illustration of the scope of applicability of our algorithm, we consider a particle moving in two-dimensional lattice with time and show that the algorithm is capable of detecting the coordinates of the particle as it moves with time. Further, to properly connect with the idea of integrating our QWSA with state-of-the-art quantum hardware, we construct an equivalent quantum circuit that can implement the algorithm. The rest of the paper is organized as follows:

\section{Quantum Walk Search Algorithm}\label{sec:quantum walk search}

Let $G = (V,E)$ be a finite $d$-regular graph, where $V$ is a set of vertices (nodes), $E$ is the set of edges connecting the nodes, and $N = |V|$ is the number of vertices. The labels of the vertices are $0$ to $N-1$, and the labels of the edges are $0$ to $d-1$. A discrete-time quantum walk on the graph $G$ generates a unitary evolution operator in the Hilbert space $\mathcal{H}_U=\mathcal{H}_{\text{pos}}\otimes \mathcal{H}_{\text{coin}}$: the position space $\mathcal{H}_\text{pos}$ and the coin space $\mathcal{H}_\text{coin}$. $\mathcal{H}_\text{pos}$ is spanned by \{$|v\rangle $: $0 \leq v \leq N-1$\}, while $\mathcal{H}_\text{coin}$ is spanned by $\{|a\rangle : 0\leq a\leq d-1\}$  represents the internal states (often called ``coin states") associated with each node. At any time $t$, the state can be represented by 
\begin{equation}
	|\Psi(t)\rangle = \sum_{a,v} \phi_{a,v}(t) |a,v\rangle \,.
\end{equation}
Each step of the DTQW is generated by a unitary operator consisting of coin operation $C$ on the internal degrees of freedom followed by a conditional position shift operation $S$ on the configuration space. Therefore, the state at time $t$ and $(t+\tau)$ (where $\tau$ is the time required to implement one step of the walk) satisfies the relation, 
\begin{equation}
	|\Psi(t+\tau)\rangle =U|\Psi(t)\rangle= S(C\otimes I) |\Psi(t)\rangle\,,
\end{equation}
whereby imposing the operator form of the evolution operator $U=S(C\otimes I)$. With this information, we are set to address the QWSA. Consider that the walker starts from an initial state, which is a uniform superposition of all states over internal and external degrees of freedom,
\begin{equation}\label{initial_state}
	|\Psi(0)\rangle=|\psi_c^{(d)}\rangle\otimes \frac{1}{N}\sum |v\rangle\,, \ \ \text{where}\ \ |\psi_c^{(d)}\rangle = \frac{1}{\sqrt{d}}\sum_{a=0}^{d-1}|a\rangle \,.
\end{equation}
where $|\psi_c^{(d)}\rangle$ is uniform superposition state in coin space. The idea behind the QWSA is starting with \eqref{initial_state}, can we define a unitary operator that localizes the state to a certain point (say $|v_0\rangle$) on the grid. This operation is mathematically represented as
\begin{equation}
	|\Psi(t)\rangle=\left(U'\right)^t |\Psi(0)\rangle\,,\ \ P_v(t)=\left|\langle v|\Psi(t)\rangle\right|^2_{\text{max}\ @\ v=v_0}\,.
\end{equation}
The mathematical equation above states simply that after $t$ such operations of a unitary operator $U'$, the wave function localizes at the point $|v_0\rangle$ where the time $t$ is related to the size of the grid and the marked node configuration. The probability of success is the maximal probability for locating the node $|v_0\rangle$ and is related to the number of times $t$, the $U'$ operator has been applied on the initial state. For a single marked node, the operator $U'$ is related to the unitary operator for the DTQW $U=S\cdot (C\otimes I)$ by the relation,
\begin{equation}\label{eq:modified_evol}
	U'=U\cdot R\,,
\end{equation}
where $R$ is called "Search Oracle" and contains the information about the marked node(s). In essence, it is a phase shift operator that reverses the phase of all but one node ({\it i.e.} the marked node) by $e^{i\pi}$. For a single marked node, the Search Oracle has a simple functional form\,\cite{PhysRevA.103.062202,portugal_quantum_2013}
\begin{equation}\label{eq:search oracle}
	R=I-2|\psi_c^{(d)}\rangle\langle\psi_c^{(d)}|\otimes |v_0\rangle\langle v_0|\,.
\end{equation}
Without the coin state, this form coincides with the Grover Search Oracle\cite{kayeIntroductionQuantumComputing2006a}

\begin{equation}\label{eq:grover_oracle}
	U_f = I - 2|v_0\rangle \langle v_0|\,.
\end{equation}
The search oracle $R$ can be easily generalized for multiple marked nodes, 
\begin{equation}\label{eq:search_oracle_multiple}
	R = I - 2\ |\psi_c^{(d)}\rangle \langle \psi_c^{(d)}|\otimes \sum_{v\in M} |v\rangle \langle v|
\end{equation}
where $M$ is set of marked multiple marked node. To illustrate the concrete structure of the algorithm, we will consider a finite two-dimensional lattice with open and periodic boundary conditions.

\subsection{Finite two-dimensional lattice}\label{finite two-dimensional lattice}

Consider the quantum-walk search algorithm in the $\sqrt{N}\times \sqrt{N}$ square lattice. We will consider both open and periodic boundary conditions. The quantum state spanned by $\{|i,j\rangle \otimes |x,y\rangle: i,j\in [0,1] \ \&\ 0\leq x,y\leq \sqrt{N}-1\} $ where $|i,j\rangle$ represents coin state, and $|x,y\rangle$ represents position state. The shift operator $S$ is the flip-flop shift operator given by 

\begin{equation}
	S|i,j\rangle \otimes |x,y\rangle =|1-i,1-j\rangle \otimes |x+(-1)^i \delta_{ij},y+(-1)^j (1-\delta_{ij})\rangle\,. 
\end{equation}
The flip-flop shift operator invert the coin state as it shift the position state. This inversion in the con state is important for speeding up search algorithms on the two-dimensional lattice\cite{PhysRevA.103.062202,portugal_quantum_2013}. In explicit notation, we can write coin state as 

\begin{equation}
	|\uparrow\rangle = |00\rangle =\begin{bmatrix}
		1 \\
		0 \\
		0 \\
		0 
	\end{bmatrix},\  |\downarrow\rangle= |11\rangle =\begin{bmatrix}
		0 \\
		0 \\
		0 \\
		1 
	\end{bmatrix},\  |\leftarrow\rangle= |01\rangle = \begin{bmatrix}
		0 \\
		1 \\
		0 \\
		0 
	\end{bmatrix},\  |\rightarrow\rangle= |10\rangle = \begin{bmatrix}
		0 \\
		0 \\
		1 \\
		0 
	\end{bmatrix}\,.
\end{equation}

%

\begin{figure}
	\begin{minipage}[c]{0.70\textwidth}
		\includegraphics[width=\textwidth]{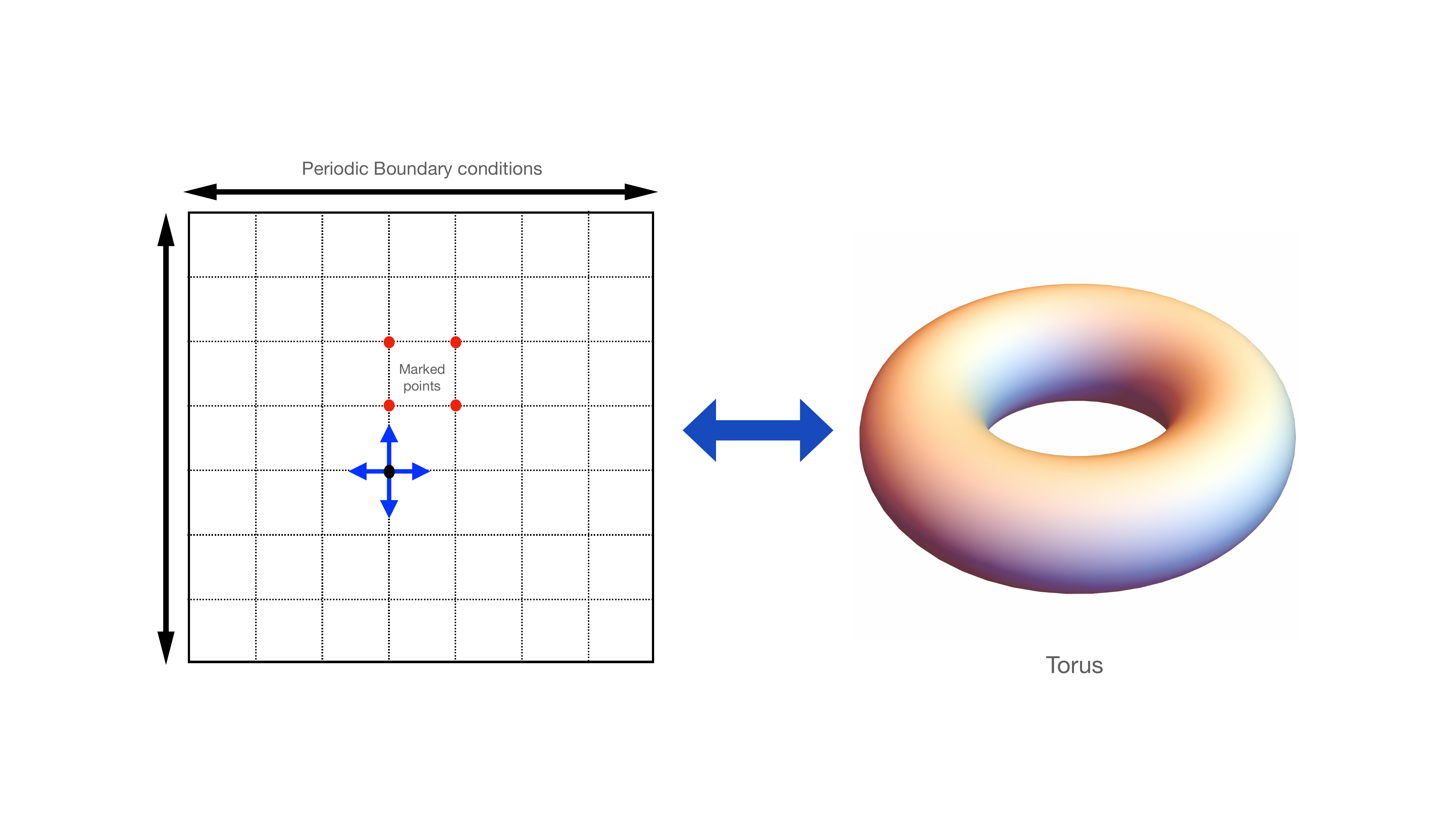}
	\end{minipage}\hfill
	\begin{minipage}[c]{0.3\textwidth}
	\caption{Two-dimensional lattice with double periodic boundary condition is topologically equivalent to a torus.}
\label{fig:gridpbc}
	\end{minipage}
\end{figure}

\begin{figure}
	\begin{minipage}[c]{0.60\textwidth}
		\centering
		\begin{tikzpicture}[scale=0.70] 
		\draw[step=1cm] (-3,-3.0) grid (3.0,3.0);
		\foreach \i in {-3,...,3}
		{
			\foreach \j in {-3,...,3}
			{
					\draw node[draw,circle,fill=white] at (\i,\j) {};
			}
		}

		\node [draw,circle,minimum size=.15cm] (r) at (0,3) {} ;
		\path (r) edge[ out=140, in=50,loop above,
		, looseness=0.8, loop
		, distance=1cm, ->]
		node[above=1pt] {} (r);
		
		\node [draw,circle,minimum size=.15cm] (r) at (0,-3) {} ;
		\path (r) edge[out=140, in=50,loop below ,looseness=0.8, distance=1cm, ->] node[above=3pt] {} (r);
		
		\node [draw,circle,minimum size=.15cm] (r) at (-3,-3) {} ;
		\path (r) edge[out=140, in=50,loop below ,looseness=0.8, distance=1cm, ->] node[above=3pt] {} (r);
		
		\node [draw,circle,minimum size=.15cm] (r) at (-3,-3) {} ;
		\path (r) edge[out=140, in=50,loop left ,looseness=0.8, distance=1cm, ->] node[above=3pt] {} (r);
		
		\node [draw,circle,minimum size=.15cm] (r) at (3,-3) {} ;
		\path (r) edge[out=140, in=50,loop below ,looseness=0.8, distance=1cm, ->] node[above=3pt] {} (r);
		
		\node [draw,circle,minimum size=.15cm] (r) at (3,-3) {} ;
		\path (r) edge[out=140, in=50,loop right ,looseness=0.8, distance=1cm, ->] node[above=3pt] {} (r);
		
		\node [draw,circle,minimum size=.15cm] (r) at (-3,0) {} ;
		\path (r) edge[out=140, in=50,loop left ,looseness=0.8, distance=1cm, ->] node[above=3pt] {} (r);
		
		\node [draw,circle,minimum size=.15cm] (r) at (-3,3) {} ;
		\path (r) edge[out=140, in=50,loop left ,looseness=0.8, distance=1cm, ->] node[above=3pt] {} (r);
		
		\node [draw,circle,minimum size=.15cm] (r) at (-3,3) {} ;
		\path (r) edge[out=140, in=50,loop above ,looseness=0.8, distance=1cm, ->] node[above=3pt] {} (r);
		
		\node [draw,circle,minimum size=.15cm] (r) at (3,0) {} ;
		\path (r) edge[out=140, in=50,loop right ,looseness=0.8, distance=1cm, ->] node[above=3pt] {} (r);
		
		\node [draw,circle,minimum size=.15cm] (r) at (3,3) {} ;
		\path (r) edge[out=140, in=50,loop right ,looseness=0.8, distance=1cm, ->] node[above=3pt] {} (r);
		
		\node [draw,circle,minimum size=.15cm] (r) at (3,3) {} ;
		\path (r) edge[out=140, in=50,loop above ,looseness=0.8, distance=1cm, ->] node[above=3pt] {} (r);

		\node [draw,circle,minimum size=.15cm] (r) at (0,0) {};
		\node [draw,circle,minimum size=.15cm] (r1) at (1,0) {};
		\node [draw,circle,minimum size=.15cm] (r2) at (-1,0) {};
		\node [draw,circle,minimum size=.15cm] (r3) at (0,1) {};
		\node [draw,circle,minimum size=.15cm] (r4) at (0,-1) {};
		
		\path[->]
		(r) edge [bend left=20] (r1)
		(r) edge [bend left=20] (r2)
		(r) edge [bend left=20] (r3)
		(r) edge [bend left=20] (r4);
		
	\end{tikzpicture}
	\end{minipage}\hfill
	\begin{minipage}[c]{0.4\textwidth}
	\caption{The structure of the shift operator in the case of open boundary conditions is different in the interior and exterior of the grid. The figure shows the self-loop at the boundary point of the lattice which ensures the unitarity of the shift operator.}
\label{fig:gridobc}
	\end{minipage}
\end{figure}
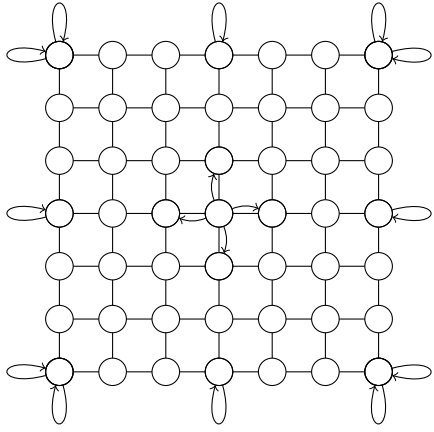

\noindent In this notation, the flip-flop ship operator for the periodic boundary condition given by 

\begin{equation}
	\begin{split}
		S_c^{(2)} &= |\downarrow\rangle  \langle \uparrow | \otimes \sum_{x,y} |x,y+1\rangle \langle x,y| +  |\uparrow\rangle  \langle \downarrow | \otimes \sum_{x,y} |x,y-1\rangle \langle x,y| \\
		&   \qquad \qquad +  |\leftarrow\rangle  \langle \rightarrow | \otimes \sum_{x,y} |x+1,y\rangle \langle x,y| +  |\rightarrow\rangle  \langle \leftarrow | \otimes \sum_{x,y} |x-1,y\rangle \langle x,y|
	\end{split}
\end{equation}
where superscript in $S_c^{(2)}$ is to emphasize that there are two directions of motion. In periodic boundary condition, the position state obey the cyclic property $|x+\sqrt{N}\rangle = |x\rangle$ and $|y+\sqrt{N} \rangle = |y\rangle $. Therefore, the geometry of two-dimensional lattice with periodic boundary condition is equivalent to torus as shown in Fig.~\ref{fig:gridpbc}. For open boundary condition, the shift operator is defined as the sum of interior term (see Fig.~\ref{fig:gridobc})

\begin{equation}
	\begin{split}
		S_\text{int}^{(2)} &= |\downarrow\rangle  \langle \uparrow | \otimes \sum_{x,y<\sqrt{N}-1} |x,y+1\rangle \langle x,y| +  |\uparrow\rangle  \langle \downarrow | \otimes \sum_{x,y>0} |x,y-1\rangle \langle x,y| \\
		&  \qquad+  |\leftarrow\rangle  \langle \rightarrow | \otimes \sum_{x<\sqrt{N}-1,y} |x+1,y\rangle \langle x,y| +  |\rightarrow\rangle  \langle \leftarrow | \otimes \sum_{x>0,y} |x-1,y\rangle \langle x,y| 
	\end{split}
\end{equation}
and boundary (exterior)  
\begin{equation}
	\begin{split}
		S_\text{ext}^{(2)} &= |\uparrow\rangle  \langle \uparrow | \otimes \sum_{x,y=\sqrt{N}-1} |x,y\rangle \langle x,y| +  |\downarrow\rangle  \langle \downarrow | \otimes \sum_{x,y=0} |x,y\rangle \langle x,y| \\
		&  \qquad +  |\rightarrow\rangle  \langle \rightarrow | \otimes \sum_{x=\sqrt{N}-1,y} |x,y\rangle \langle x,y| +  |\leftarrow\rangle  \langle \leftarrow | \otimes \sum_{x=0,y} |x,y\rangle \langle x,y| 
	\end{split}
\end{equation}
as $S_\mathrm{o}^{(2)} = S_\text{int}^{(2)} + S_\text{ext}^{(2)}$. The explicity form of $S_\text{int}^{(2)}$ and $S_\text{ext}^{(2)}$ is chosen so that the shift operator $S_\mathrm{o}^{(2)}$ is unitary i.e. $S_\mathrm{o}^{(2)}(S^{(2)}_\mathrm{o})^\dagger = (S^{(2)}_\mathrm{o})^\dagger S_\mathrm{o}^{(2)}= I$ (see supplementary material). The coin operator in the quantum-walk search is chosen to be the Grover diffusion operator\,\cite{Tregenna_2003,PhysRevA.77.062331} 

\begin{equation}
	G^{(2)} = 2|\psi_c^{(2)} \rangle \langle\psi_c^{(2)} | - I = \frac{1}{2} \begin{bmatrix}
		- 1 & 1 & 1 & 1 \\
		1 & -1 & 1 & 1 \\
		1 & 1 & -1 & 1 \\
		1 & 1 & 1 & -1 \\
	\end{bmatrix}\,.
\end{equation}
The state is initialized as an equal superposition in coin and position space, given by 

\begin{equation}
	|\Psi(0)\rangle = \frac{1}{2}\sum_{i,j}|i,j\rangle \otimes\frac{1}{N} \sum_{x,y} |x,y\rangle =  |\chi\rangle \otimes   \frac{1}{N} \sum_{x,y} |x,y\rangle 
\end{equation} 
The modified unitary operator in Eq.~\eqref{eq:modified_evol} is applied optimal time $t_\text{op}$ times to initial state 
\begin{equation}
	|\Psi(t_\text{op})\rangle  = (U')^{t_\text{op}}|\Psi(0)\rangle 
\end{equation}
which amplifies the probability amplitude of target states. For the two-dimensional lattice, the complexity of the algorithm is given by $\mathcal{O}(\ln N)$. The running time is $t_\text{op}= \mathcal{O}(\sqrt{N\ln N})$ and the success probability is $p_\text{succ} = \mathcal{O}(1/\ln N)$ \cite{portugal_quantum_2013}. 

\section{Results}
\subsection{QWSA for ordered marked nodes}\label{sec:quantum walk search for ordered marked nodes}

In previous literature \cite{xu_robust_2022,wong_exceptional_2017,j_spatial_2018,li_generalized_2020,glos_upperbounds_2021,PhysRevA.103.062202}, quantum-walk search with multiple marked points has been extensively studied.  In a general QWSA, multiple marked points can indeed exist on the graph, and there is no inherent (chronological) ordering associated with these marked points. The algorithm's objective is to efficiently locate any one of these marked points without any preference for their order. In this section, we consider that there is an  additional (and preferably a chronological) ordering associated with marked points, and devise a refined algorithm that addresses this ordering. More generally, our refined algorithm will address the case where we have multiple marked points belonging to different categories and we will be searching for the point along with its category.\\

%
%
%

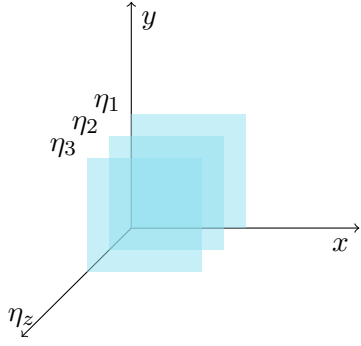
\begin{figure}
	\begin{minipage}[c]{0.6\textwidth}
		\centering

	\begin{tikzpicture}[scale=1.5]
		\draw[->] (0,0,0) -- (2,0,0) node[anchor=north east]{$x$};
		\draw[->] (0,0,0) -- (0,2,0) node[anchor=north west]{$y$};
		\draw[->] (0,0,0) -- (0,0,2.5) node[anchor=south]{$\eta_z$};
		\fill[mycolor, opacity=0.5] (0,0,0) rectangle (1,1,0);
		\fill[mycolor, opacity=0.5] (0,0,0.5) rectangle (1,1,0.5);
		\fill[mycolor, opacity=0.5] (0,0,1) rectangle (1,1,1);
		\node[anchor=east] at (0,1.1,0) {$\eta_1$};
		\node[anchor=east] at (0,1.1,0.5) {$\eta_2$};
		\node[anchor=east] at (0,1.1,1) {$\eta_3$};
		
	\end{tikzpicture}
	\end{minipage}\hfill
	\begin{minipage}[c]{0.4\textwidth}
	\caption{Schematic diagram of the structure of Hilbert space for quantum-walk search for ordered marked nodes. }
\label{fig:schematiclayerhanddrawn}
	\end{minipage}
\end{figure}

\noindent For starters, consider a total of $m$ different categories $\eta_z$ where $z = 0,1,2,\ldots ,m-1$. Further, we assume that a particular category $\eta_z$ has a unique marked point. To represent this system, we introduce additional label states $|\eta_z\rangle$ with $0\leq z\leq m-1$, adding an extra dimension to the Hilbert space. We consider a finite two-dimensional lattice, although our method can be easily generalized to arbitrary graphs and dimensions. Fig.~\ref{fig:schematiclayerhanddrawn} shows the schematic diagram of Hilbert space for the search algorithm, which shows replicated layers of the $2d$ lattice representing different categories $\eta_z$. Depending on whether the motion between different layers (categories) is allowed or not, we have two different scenarios: Static labeling and Dynamic labeling.

\subsubsection{Static Labelling}
Consider the case where the walker is not allowed to move between the layers. The Hilbert space is spanned by basis $\{|i,j\rangle \otimes |x,y\rangle \otimes |\eta_z\rangle: i,j \in [0,1]\ \&\ 0\leq x,y\leq N-1\ \&\ 0\leq z\leq m-1\}$. The oracle can be written as

\begin{equation}\label{eq:multilayered search oracle}
	R = I - 2|\psi^{(2)}_c\rangle \langle \psi^{(2)}_c|\otimes \sum_{z} \sum_{\mathcal{X}\in M_z} |x,y\rangle \langle x,y|\otimes |\eta_z\rangle \langle \eta_z| 
\end{equation}
where $M_z$ is set of marked nodes in layer $z$, and $\mathcal{X} \equiv (x,y)$ to unclutter the notations. Since the walker doesn't move between the layers, the shift and coin operator are the same as the unordered marked case as described in Sec.~\ref{sec:quantum walk search}. Note that the Hilbert space for this case is reducible to a direct sum of Hilbert spaces associated with different layers. Therefore, the algorithm boils down to a set of independent reduced QWSA on each layer. Following the reducibility of the Hilbert space, we can write  the evolution operator as a direct sum of evolution operators of individual layers
\begin{equation}
	U' = U\cdot R= \bigoplus_{z} U'_z = \bigoplus_{z} \left(U_z \cdot R_z\right)\,,
\end{equation}
where $U_z = S_z(C_z\otimes I)$ and 
\begin{equation}
	R_z = I - 2 |\psi_c^{(2)} \rangle \langle \psi_c^{(2)} |\otimes \sum_{\mathcal{X}\in M_z} |x,y\rangle \langle x,y|\,.
\end{equation}
The initial state is an equal superposition in coin, position, and label space, given by 
\begin{equation}
	|\Psi(0)\rangle = \frac{1}{2}\sum_{i,j} |i,j\rangle \otimes \frac{1}{N \sqrt{m}}\sum_{x,y,z} |x,y\rangle \otimes |\eta_z\rangle
\end{equation}
The modified operator $U'$ is applied $t_\text{op}$ times to the state, which amplifies the marked points and their associated labels. To illustrate the algorithm, we perform a numerical simulation of the algorithm on $16\times16$ grid under both open and periodic boundary conditions. The simulation includes four marked nodes at position $\{(6,8),(8,9),(12,5),(15,5)\}$, which necessitates four layers or label states.

\begin{figure}[ht]
	\centering
	\begin{subfigure}[b]{0.45\textwidth}
		\centering
		\includegraphics[scale = 0.45]{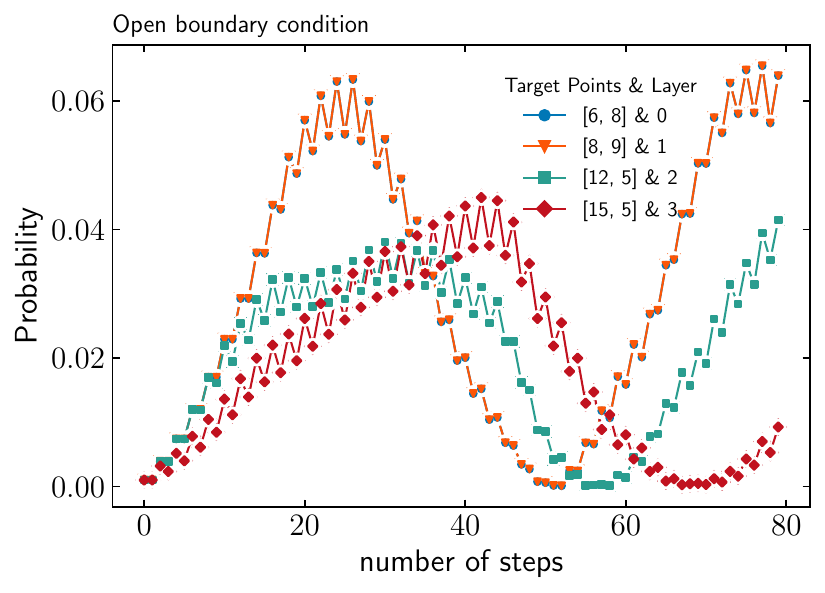}
		\caption{$2d$ open grid}
	\end{subfigure}
	\hfill
	\begin{subfigure}[b]{0.45\textwidth}
		\centering
		\includegraphics[scale = 0.45]{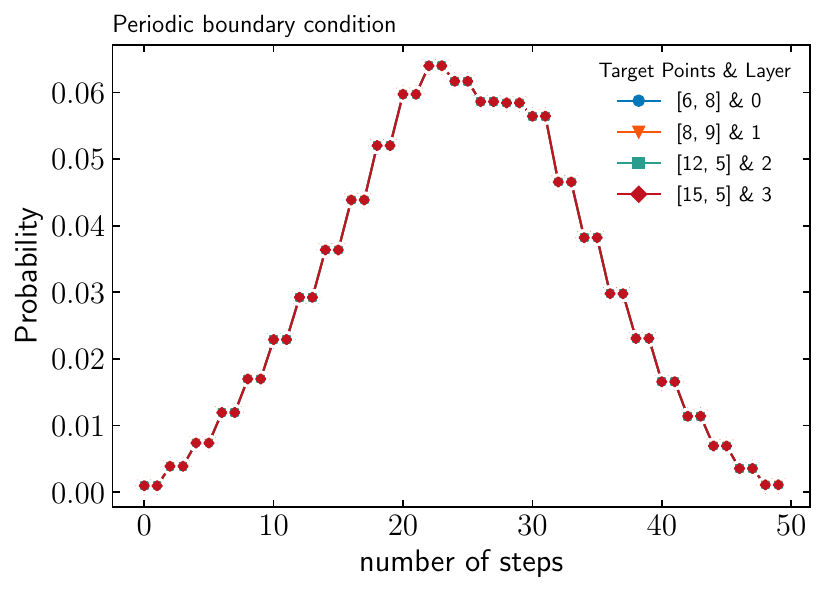}
		\caption{Torus}
	\end{subfigure}
	\caption{Amplification of marked nodes with steps for static labelling in the case of \textbf{(a)} open grid and \textbf{(b)} torus. In case of open grid, the probability of marked nodes $[6,8]$ and $[8,9]$ coincides, while in case of torus, the probability of all marked nodes coincide with each other. }
	\label{fig:single_layer_prob}
\end{figure}

\begin{figure}[ht]
	\centering
	\includegraphics[scale = 0.55]{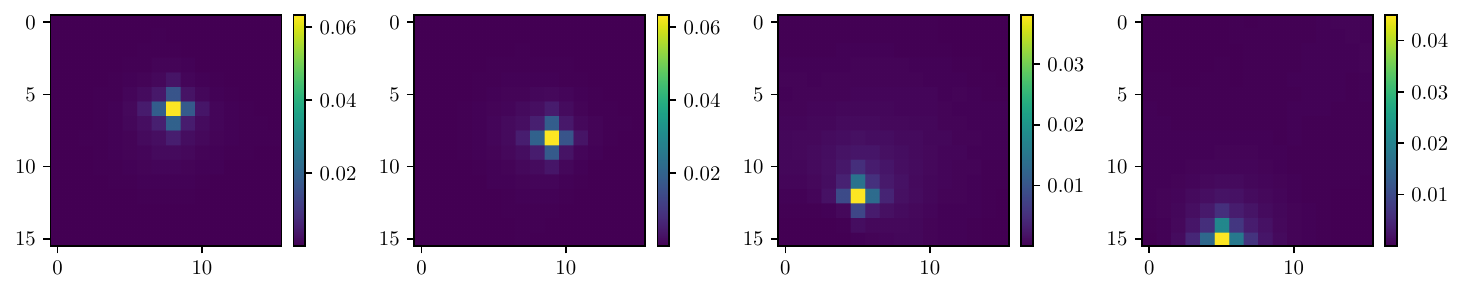}
	\includegraphics[scale = 0.55]{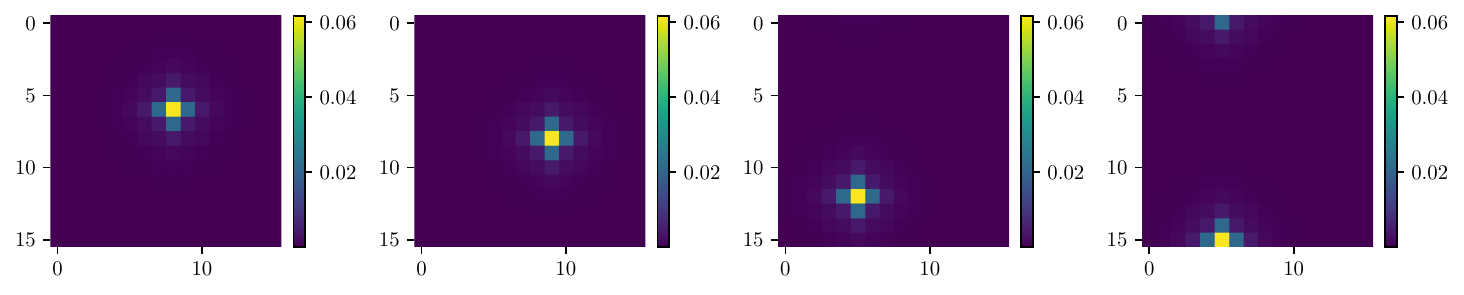}
	\caption{Probability distribution at $t_\text{op}$ step for different layers with a single marked point in each layer found using quantum-search algorithm with static labelling \textbf{Top}: Open boundary condition \textbf{Below} Periodic boundary condition}
	\label{fig:single_layer}
\end{figure}

Figure~\ref{fig:single_layer_prob} shows the probability of finding a labeled marked node as a function of the number of steps taken by the algorithm. Figure~\ref{fig:single_layer} displays the probability distribution at the optimal time step. For the periodic boundary condition, we observe that the probabilities of finding each marked node coincide. This results from the translational symmetry inherent in the toroidal geometry of the lattice. In contrast, under the open boundary condition, the probability of finding a marked node generally depends on its location due to boundary effects. Increasing the system size would lead us to anticipate the disappearance of boundary effects. Consequently, we expect the two probability distributions to converge as the number of nodes $N$ approaches infinity.

As a consequence of separability of algorithm to independent QWSA on each layer, the optimal time $t_\text{op}$ is the same as that for a conentional QWSA. Although, since the probability weight of wavefunction on each layer is scaled down by the factor of number of layers, as a result the success probability $p_\text{succ}$ is also scaled down by the number of the same factor. Therefore, the success probability associated with each layer is $\mathcal{O}(1/(m\ln N))$.

\subsubsection{Dynamic Labelling}
For the dynamical labeling, where we allow for inter-layer transition, the coin space includes an additional direction to facilitate such motion (along the direction of the labels). The corresponding Hilbert space is spanned by a basis $\{|i,j,k\rangle \otimes |x,y\rangle \otimes |\eta_l\rangle: i,j,k \in [0,1]\ \&\ 0\leq x,y\leq N-1\ \&\ 0\leq k\leq m-1\}$ which has dimension $8mN^2$. The Grover diffusion operator $G_3$ is elevated to a three-qubit operator. There can be further analogous extensions to open and periodic boundary conditions along the label direction, but we will stick to open boundary along the label direction. Similarly, the shift operator becomes, 
\begin{equation}
	S|i,j,k\rangle \otimes |x,y,\eta_z\rangle = |1-i,1-j,1-k\rangle \otimes |x+(-1)^i(1-\delta_{ij}),y+(-1)^i\delta_{ij},\eta_{z+(-1)^k}\rangle   
\end{equation}
The modified unitary evolution for the search given by 
\begin{equation}
	U' = U\cdot R = S(G_3\otimes I)\cdot R
\end{equation}
where the search oracle is given by 

\begin{equation}
	R = I - 2|\psi^{(3)}_c\rangle \langle \psi^{(3)}_c|\otimes \sum_{z} \sum_{\mathcal{X}\in M_z} |x,y\rangle \langle x,y|\otimes |\eta_z\rangle \langle \eta_z|\,.
\end{equation}
The initial state is a uniform superposition in Hilbert space. The modified operator $U'$ is applied $t_\text{op}$ times on this state, which amplifies the marked point along with labels.

\begin{figure}[ht]
	\centering
	\begin{subfigure}[b]{0.45\textwidth}
		\centering
		\includegraphics[scale = 0.45]{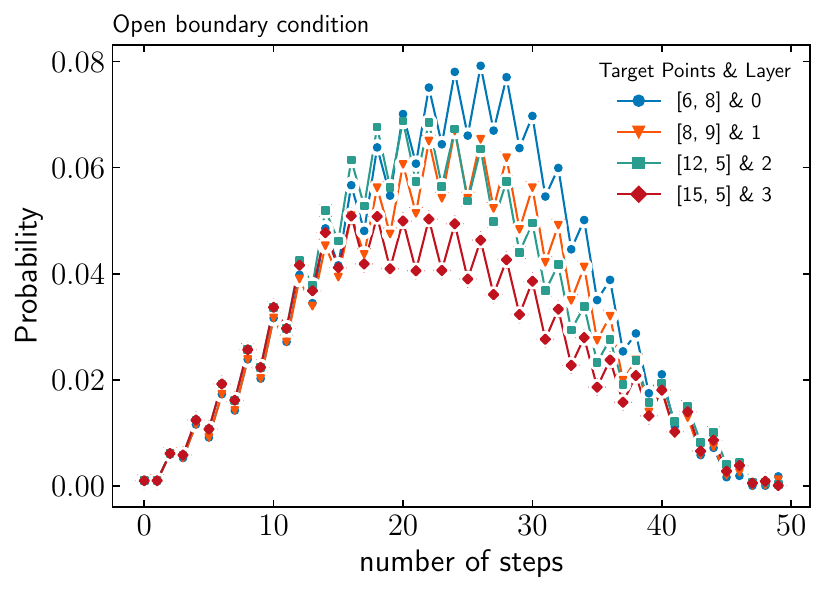}
		\caption{$2d$ open grid}
	\end{subfigure}
	\hfill
	\begin{subfigure}[b]{0.45\textwidth}
		\centering
		\includegraphics[scale = 0.45]{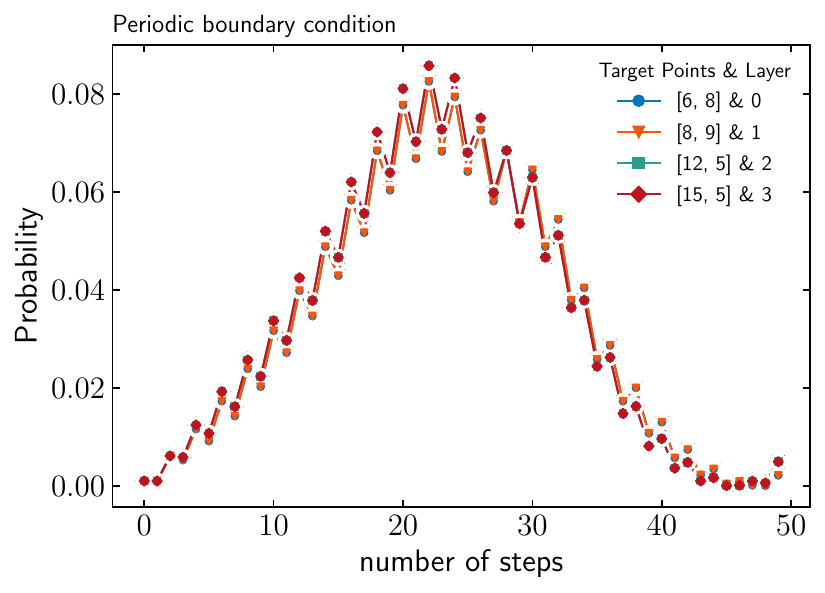}
		\caption{Torus}
	\end{subfigure}
	\caption{Amplification of marked nodes with steps for dynamic labelling in the case of \textbf{(a)} open grid and \textbf{(b)} torus.}
	\label{fig:multi_layer_prob}
\end{figure}

\begin{figure}[ht]
	\centering
	\includegraphics[scale = 0.55]{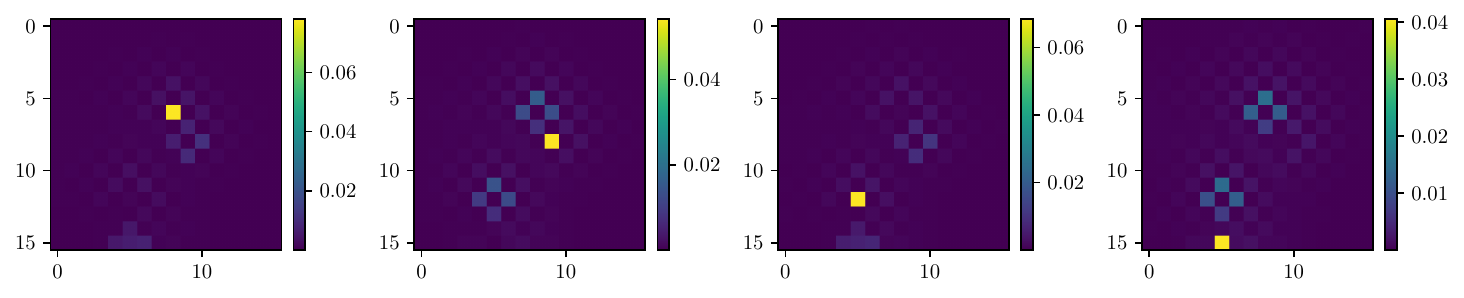}
	\includegraphics[scale = 0.55]{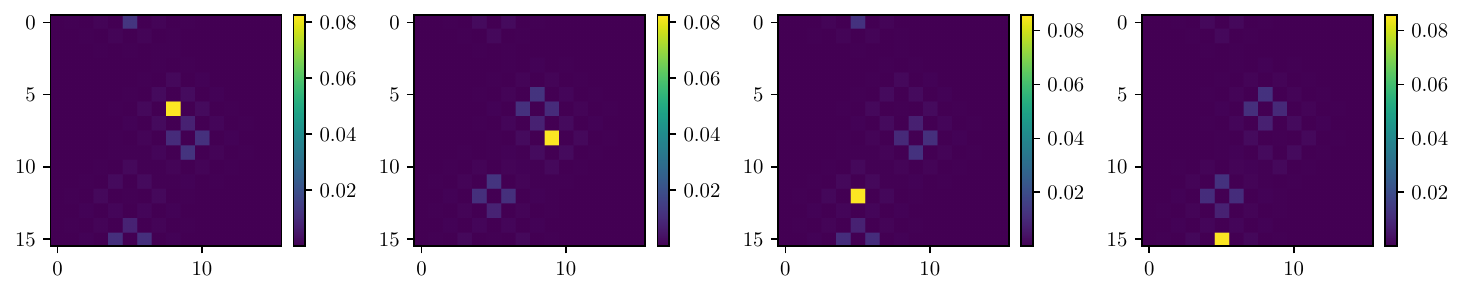}
	\caption{Probability distribution at $t_\text{op}$ step for different layers with a single marked point in each layer found using quantum-search algorithm with dynamic labelling \textbf{Top}: Open boundary condition \textbf{Below} Periodic boundary condition}
	\label{fig:multi_layer}
\end{figure}

We demonstrate the amplification due to dynamic labeling for the case of the open grid and torus with the same parameters as static case in Fig~\ref{fig:multi_layer_prob}, and probability distribution at optimal time-step in Fig.~\ref{fig:multi_layer}. Note that in this case, the torus (as in the case of static labeling) has a unique $t_\text{op}$. While the open grid performs much better than the static labeling and possesses a close to unique $t_\text{op}$ where all the marked nodes are simultaneously amplified.

\subsubsection{Scaling}\label{subsec:Scaling}

\begin{figure}[ht]
	\centering
	\begin{subfigure}[b]{0.24\textwidth}
		\centering
		\includegraphics[width = \textwidth]{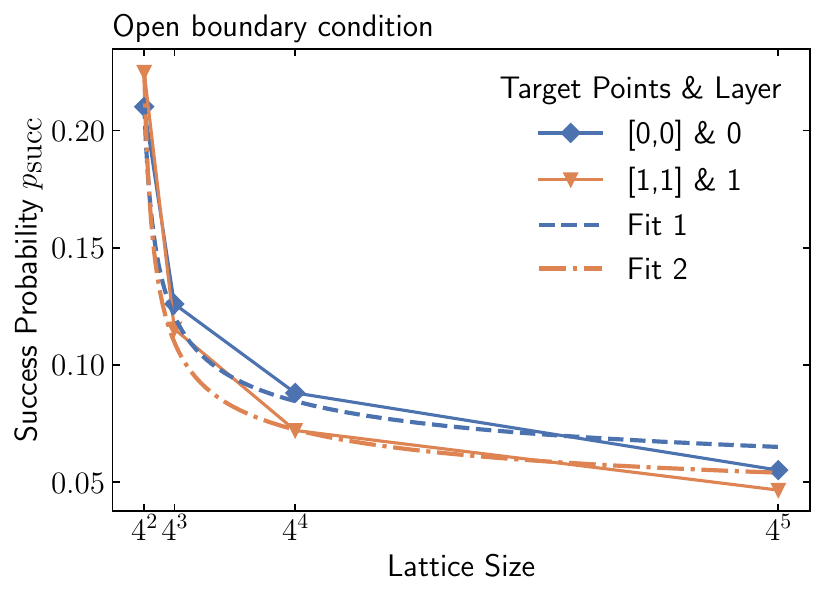}
	\end{subfigure}
	\hfill
	\begin{subfigure}[b]{0.24\textwidth}
		\centering
		\includegraphics[width = \textwidth]{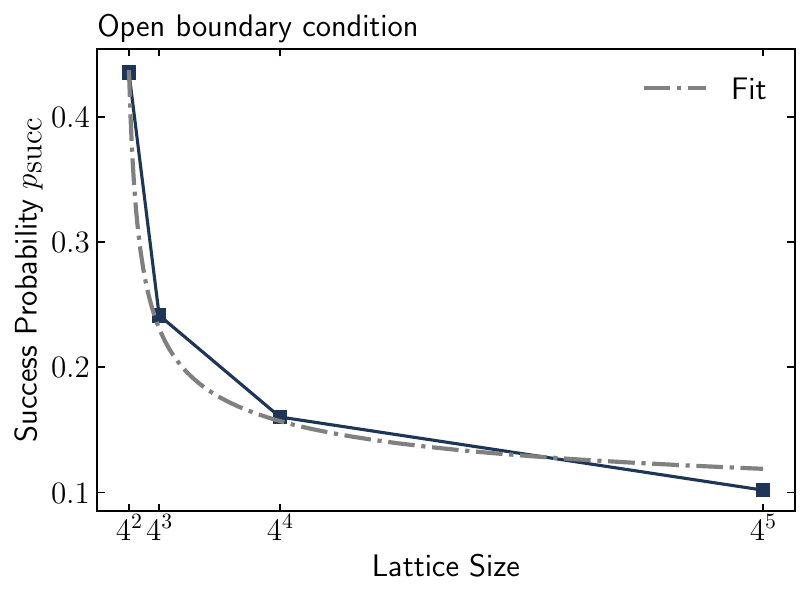}
	\end{subfigure}
	\hfill
	\begin{subfigure}[b]{0.24\textwidth}
		\centering
		\includegraphics[width = \textwidth]{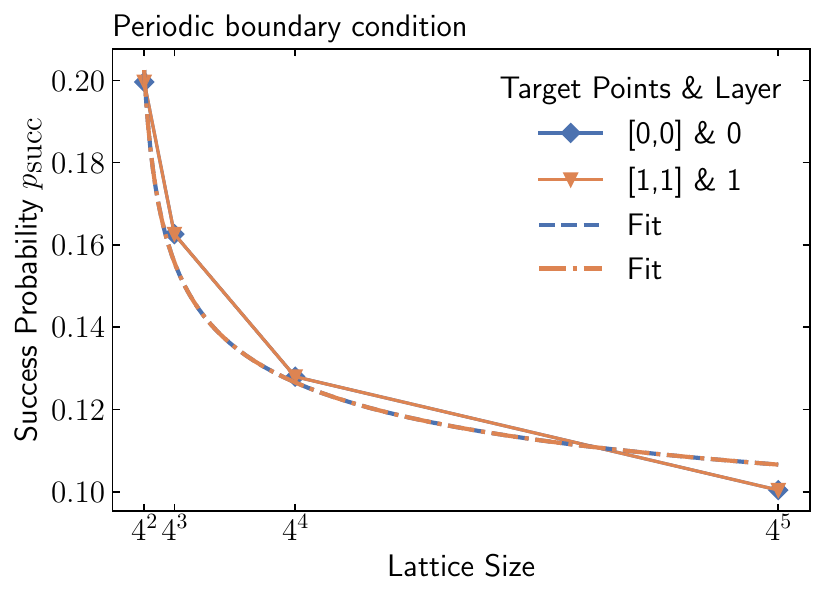}
	\end{subfigure}
	\hfill
	\begin{subfigure}[b]{0.24\textwidth}
		\centering
		\includegraphics[width = \textwidth]{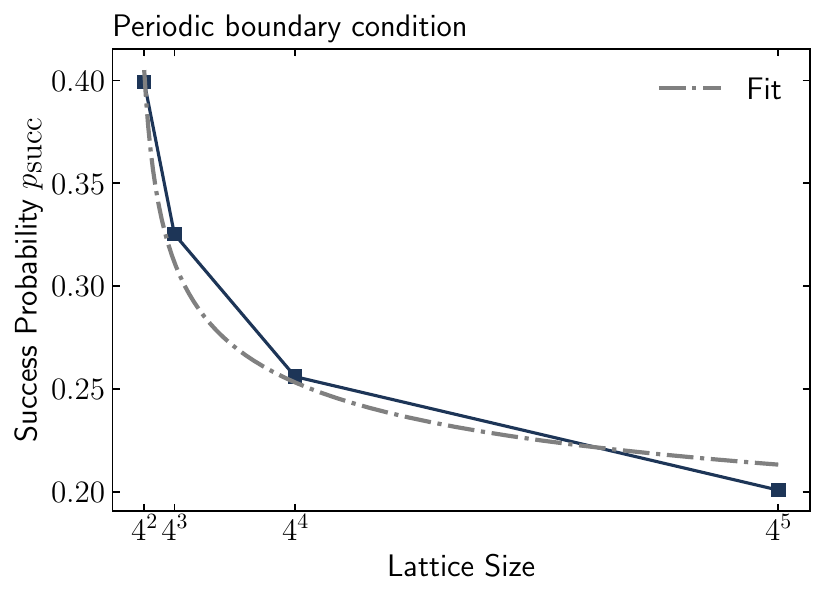}
	\end{subfigure}
	\hfill
	\begin{subfigure}[b]{0.24\textwidth}
		\centering
		\includegraphics[width = \textwidth]{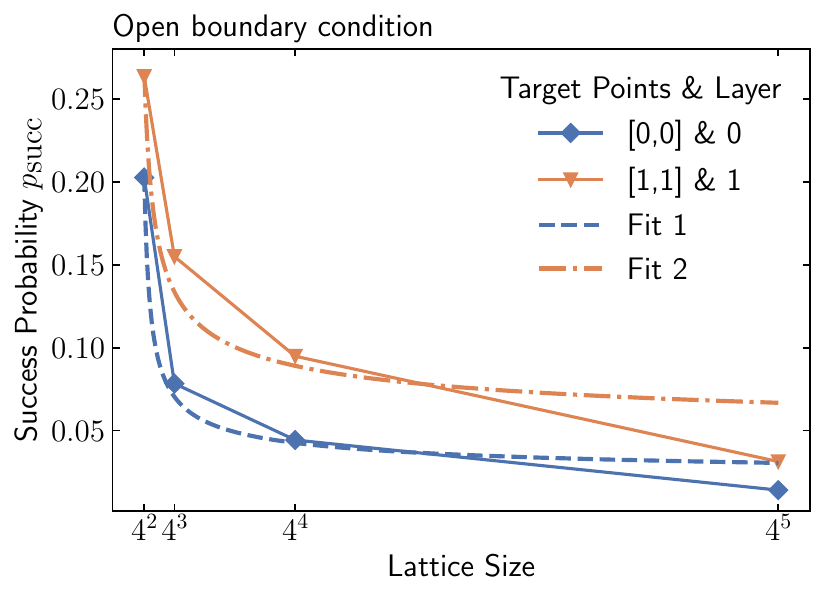}
	\end{subfigure}
	\hfill
	\begin{subfigure}[b]{0.24\textwidth}
		\centering
		\includegraphics[width = \textwidth]{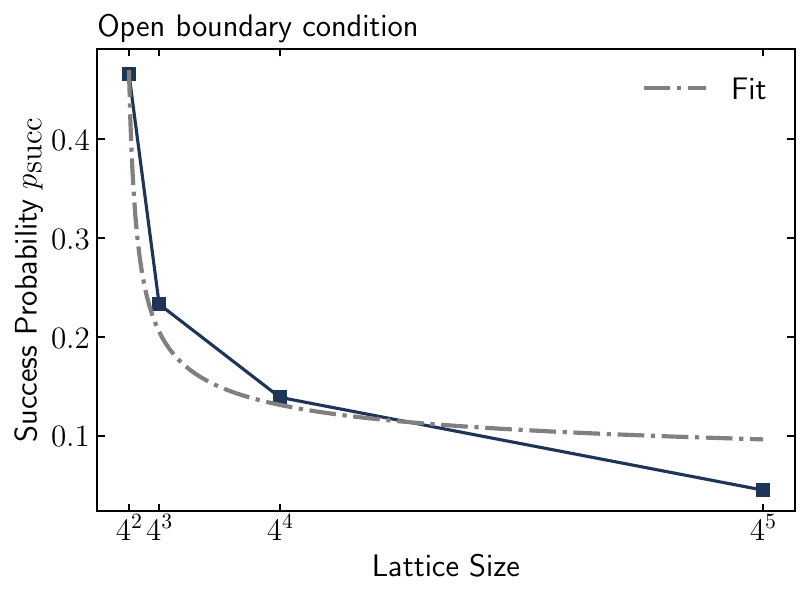}
	\end{subfigure}
	\hfill
	\begin{subfigure}[b]{0.24\textwidth}
		\centering
		\includegraphics[width = \textwidth]{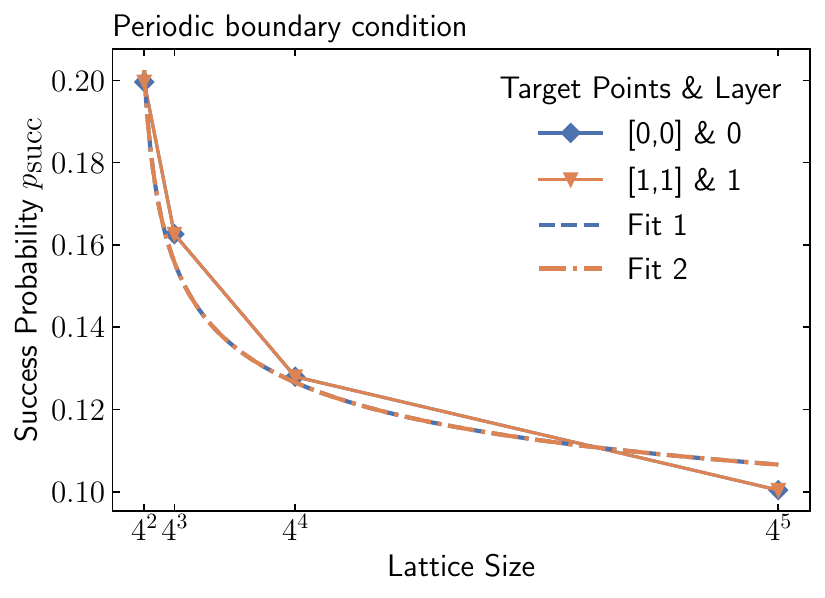}
	\end{subfigure}
	\hfill
	\begin{subfigure}[b]{0.24\textwidth}
		\centering
		\includegraphics[width = \textwidth]{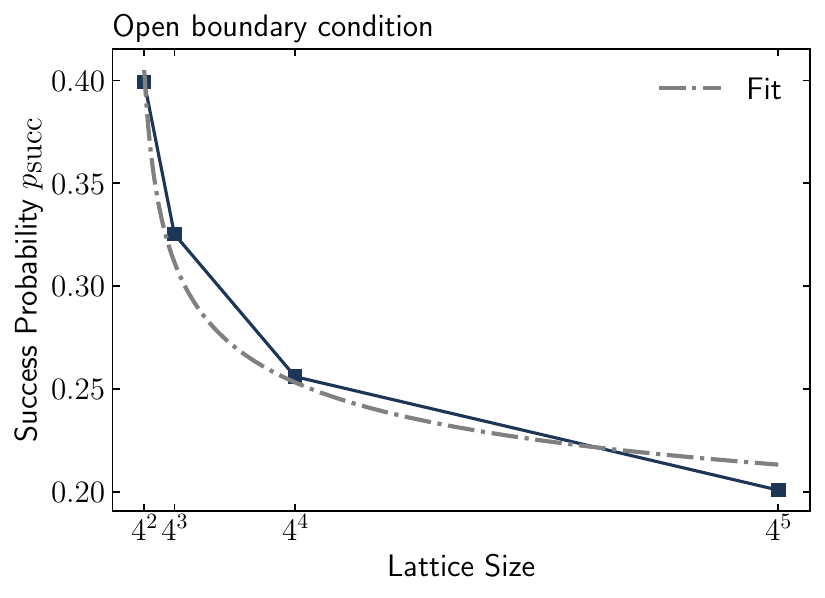}
	\end{subfigure}
	\caption{Success probability of marked points and total success probability as a function of lattice size $N$. \textbf{Top:} Static Labelling \textbf{Below:} Dynamic Labelling}
	\label{fig:scaling_static}
\end{figure}

This section investigates how the success probability of our search algorithm scales with the lattice size. We illustrate this by considering the algorithm with two marked nodes, $\{[0,0],[1,1]\}$, on lattices of varying sizes. In Figure~\ref{fig:scaling_static} presents the success probability of individual marked points and the collective success probability (sum of all individual success probabilities) for both static and dynamical cases.  We fit the curve $a/\log(bN)$, which indicating that the algorithm exhibits similar scaling behavior as the conventional QWSA in two-dimensional lattices\,\cite{portugal_quantum_2013}. Therefore, the collective sucess probability of our search algorithm scales as $\mathcal{O}(1/\ln(N))$, while the success probability of individual marked point scales as $\mathcal{O}(1/m\ln(N))$.

\subsection{Quantum tracking problem}\label{sec:quantum tracking problem}

So far, we have examined our refined QWSA for searches on $2d$ surfaces using open and periodic boundary conditions. We have established that in most cases, the algorithm works better for simultaneous amplification of multiple marked nodes at a unique time $t_\text{op}$. In simple terms, this implies that there exists a unitary operator $U$ such that $U^{t_\text{op}}$ acting on a maximally superposed initial state can maximize the probability of the marked nodes. In this section we demonstrate a practical application of the algorithm introduced in Sec.~\ref{sec:quantum walk search for ordered marked nodes} for tracking a particle moving in real-time. \\
We consider a particle moving on a $2d$ surface and the time taken by the particle to move one step is $\delta t$. Let us assume that position of particle at an instance of time $t_z = z\delta t$ with $z = 0,1,2,\ldots$  is $\mathcal{X}_z = (x_z,y_z)$. Our aim is to find the trajectory of the particle i.e. $\mathcal{X}_z$. 

A two-dimensional lattice represents the particle's configuration space and labels represent time steps. For example $\mathcal{X}_z=(x_z,y_z)$ represents the coordinates of the particle at time $t=z\delta t$. However, it would seem that associating labels with time has an obvious disadvantage in terms of resources, since the time variable continues to increase and so does the labels, hinting at a requirement of potentially infinite resource well. This in turn makes our labelling algorithm practically inapplicable owing to our limited resources. To overcome this problem, we will recycle our labels. Let's understand this in more detail. Let us define layers $l_0,l_1,l_2,\ldots$ representing the configuration space of a particle at time $0,\delta t,2\delta t,\ldots$. Furthermore, we assume that the probability amplification takes computational time ($s_i$) such that  $s_0,s_1,s_2,\ldots \ll O(T)$. In general, the maximum time of amplification $T$ is greater than the time step $\delta t$. Therefore, the information about the particle's appearance must remain in the constructed oracle for at most time $T$. Let us define $m$ as $m = [ T/\delta t]$ which is the number of steps that particles take in time $T$ which is the least number of layers required. Therefore, we can write the oracle as

\begin{equation}
	\begin{split}
		R &= I - 2|\psi_c^{(2)} \rangle \langle \psi_c^{(2)} |\otimes  \left[f(0,T) |x_0,y_0,\eta_0\rangle \langle x_0,y_0,\eta_0| + f(\delta t,T + \delta t) |x_1,y_1,\eta_1\rangle \langle x_1,y_1,\eta_1| \right. \\
		& \qquad \qquad \qquad \left. + \cdots + f(m\delta t,T + m\delta t)|x_m,y_m,\eta_m\rangle \langle x_m ,y_m,\eta_m| \right. \\
		& \qquad  \qquad \qquad \left. + f((m+1)\delta t,T + (m+1)\delta t) |x_{m+1},y_{m+1},\eta_0\rangle \langle x_{m+1},y_{m+1},\eta_0| + \cdots  \right] \\ 
		&= I - 2|\psi_c^{(2)} \rangle \langle \psi_c^{(2)} |\otimes  \left[f(0,T) |x_0,y_0,\eta_0\rangle \langle x_0,y_0,\eta_0| + f(\delta t,T + \delta t) |x_1,y_1,\eta_1\rangle \langle x_1,y_1,\eta_1| \right. \\
		& \qquad \qquad \qquad \left. + \cdots + f(T,2T)|x_m,y_m,\eta_m\rangle \langle x_m ,y_m,\eta_m| \right. \\
		&  \qquad \qquad \qquad \left. + f(T+\delta t,2T + \delta t) |x_{m+1},y_{m+1},\eta_0\rangle \langle x_{m+1},y_{m+1},\eta_0| + \cdots  \right] \\
		& = I - 2|\psi_c^{(2)} \rangle \langle \psi_c^{(2)} |\otimes \sum_{n\in \mathbb{Z}} f(n\delta t,n\delta t +T) |x_n,y_n,\eta_{n \text{ mod } m}\rangle \langle x_n,y_n,\eta_{n \text{ mod } m}|
	\end{split}
\end{equation}

where $f(x,y) = \Theta(x) - \Theta (y)$, and  $\Theta(x)$ is Heaviside step function. The shift-operator is used in accord with boundary conditions, and the coin-operator is the Grover diffusion operator $G_2$ as we are considering single-layer amplification. The time profile of probability distribution for different layers is provided in supplementary material.

\section{Methods}
\subsection{Quantum circuit implementation}\label{sec:quantum circuit implementation}

\begin{figure}
	\begin{minipage}[c]{0.6\textwidth}
		\includegraphics[width=\textwidth]{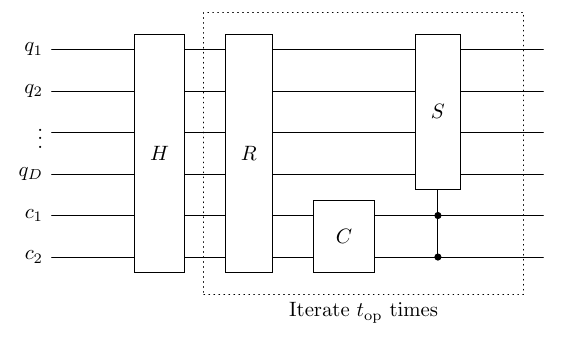}
	\end{minipage}\hfill
	\begin{minipage}[c]{0.35\textwidth}
	\caption{Schematic of the quantum circuit for quantum-walk search.}
\label{fig:qwscircuit}
	\end{minipage}
\end{figure}

In this section, we will propose quantum circuit implementation for quantum walk search for ordered marked nodes as well as quantum tracking problems that have a similar structure. In Fig.~\ref{fig:qwscircuit}, we show a schematic of a quantum circuit for quantum walk search. The qubits $q_0,q_1,\ldots ,q_D$ represents position space so that $2^D = N$, and $c_1,c_2$ represents coin space. The initial state which is a uniform superposition in position and coin space is constructed through the Hadamard operator. The modified unitary operator is then applied $t_\text{op}$ times which gives probability amplitude amplification for marked points.\\

\begin{figure}
	\begin{minipage}[c]{0.6\textwidth}
		\includegraphics[width=\textwidth]{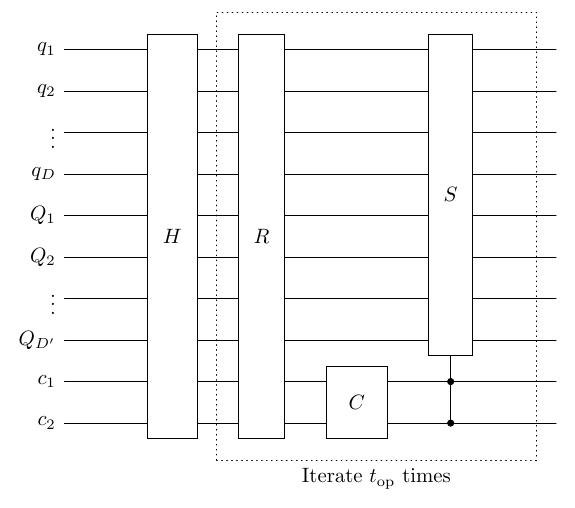}
	\end{minipage}\hfill
	\begin{minipage}[c]{0.35\textwidth}
	\caption{Schematic of the quantum circuit for quantum-walk search for ordered marked points.}
\label{fig:qwscircuit_ordered}
	\end{minipage}
\end{figure}

\noindent For our QWSA with labelled marked points, we introduce extra qubits for the layers. Fig.~\ref{fig:qwscircuit_ordered} represents a schematic of a quantum circuit for static labelling with additional $Q_1,Q_2,\ldots ,Q_{D'}$ qubits for $2^{D'} = m$ layers (or labels). The circuit for dynamic labelling is similar except we have three qubits for coin space. The equivalent circuit for quantum tracking is also similar to that for QWSA with static labelling. The specific structure of the oracle and other elements depends on configurations of marked points and  turns out to be control unitary operations. We construct the coin, shift and the oracle operator, explicitly below.

%

\subsubsection{Coin-Operator}

\begin{figure}[ht]
	\centering
	\includegraphics[scale=1]{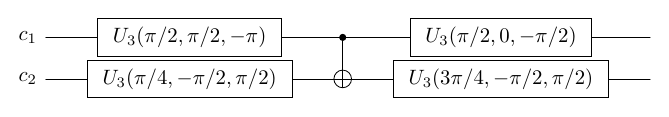}
	\caption{Implementation of Grover's diffusion operator for $2$-qubits done using Qiskit}
	\label{fig:grover_diffusion_op}
\end{figure}

The explicit implementation of coin and shift operator, and complexity in the discrete-time quantum walk has been previously done in \cite{fillion-gourdeau_algorithm_2017,huerta_alderete_quantum_2020,puengtambol_implementation_2021,bhattacharya_complexity_2023}. In QWSA, the coin operator is a Grover diffusion operator in two qubits. The optimal circuit construction for a qubit real unitary operator requires at most 2 CNOT and 12 one-qubit gates \cite{PhysRevA.69.032315}. Although, the Grover diffusion operator can be implemented with 1 CNOT and 4 one-qubits gates (See Fig.~\ref{fig:grover_diffusion_op}).

\noindent Another way to implement the Grover diffusion operator is to write it in Hadmard basis, in which, it is given by\cite{scottlecturenotes}

\begin{equation}
	A = \begin{bmatrix}
		1 & 0 & 0 & 0 \\
		0 & -1 & 0 & 0 \\
		0 & 0 & -1 & 0 \\
		0 & 0 & 0 & -1
	\end{bmatrix}
\end{equation}
Implementing the operator $A$ as a quantum circuit becomes straightforward by incorporating ancilla qubits. These ancilla qubits serve to verify whether the input comprises entirely of 0's, allowing for the inversion of the phase if it does not.

\subsubsection{Shift-Operator}

\begin{figure}[ht]
	\centering
	\begin{subfigure}[b]{0.45\textwidth}
		\centering
		\includegraphics[width=\textwidth]{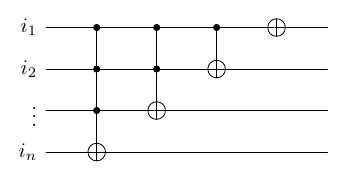}
		\caption{}
	\end{subfigure}
	\hfill
	\begin{subfigure}[b]{0.45\textwidth}
		\centering
		\includegraphics[width=\textwidth]{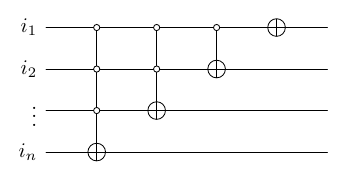}
		\caption{}
	\end{subfigure}
	\caption{Translation operator for qubits in position space. (a) $n$-qubit incrementor circuit. (b) $n$-qubit decrementor circuit.}
	\label{fig:incrementor}
\end{figure}

\begin{figure}[ht]
	\centering
	\includegraphics[scale=0.8]{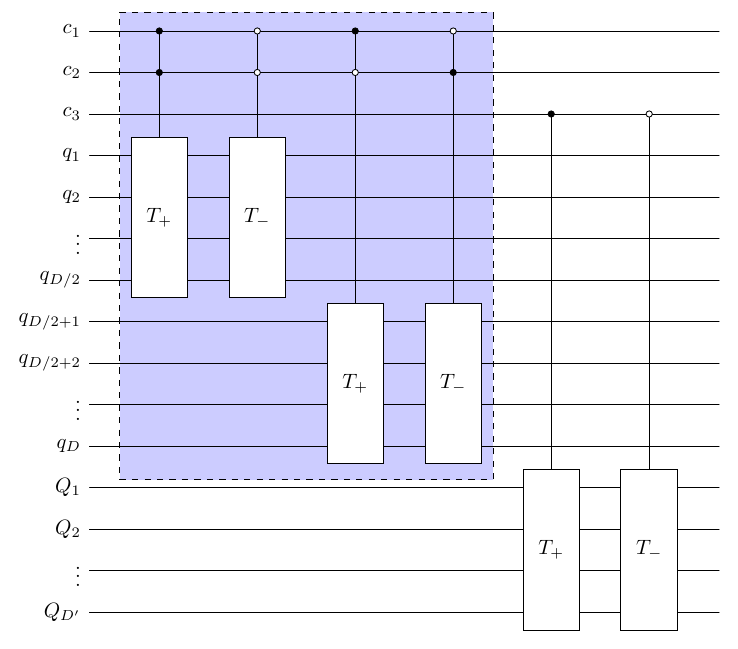}
	\caption{Quantum circuit for flip-flop shift operator where $T_+$ and $T_-$ represent translation operators (incrementor and decrementor respectively). The qubits $q_1,q_2,\ldots, q_{D/2}$ and $q_{D/2+1},q_{D/2+2},\ldots q_D$ represents position space qubits associated $x$ and $y$-direction respectively, $c_1,c_2$ are coin-space qubits, and $Q_1,Q_2,\ldots ,Q_{D'}$ represent layer states. The dotted part represents the circuit needed for static labelling.}
	\label{fig:shift_op}
\end{figure}

The flip-flop shift operator is a conditional incrementor over position space qubits. To explicitly implement this, we start with mapping computational basis associated with position $x = 0,1,2,\ldots ,\sqrt{N}-1$ into qubit basis by representing state $|x\rangle$ into its binary representation (similar for $y$ direction). As discussed in \cite{puengtambol_implementation_2021}, we can construct an incrementor circuit as shown in Fig.~\ref{fig:incrementor} using a series of multi-qubit CNOT gates. A $n$-qubit CNOT gate can be decompose into $\approx 16n$ Toffoli gates, achieving $\mathcal{O}(n)$ bound \cite{gidney_constructing_nodate}. An analogous circuit of decrementor can also be constructed using a multi-qubit CNOT gate by changing the control qubits as shown in Fig.~\ref{fig:incrementor}. We can, therefore, construct a flip-flop shift operator using a conditional operator over coin qubits as shown in Fig.~\ref{fig:shift_op}.

\noindent In case of multiple layers, the translation operator couples with qubit representing layers depending on static or dynamic labelling. As we seen in Sec.~\ref{sec:quantum walk search for ordered marked nodes}, the algorithm decouples for static labelling, therefore the shift operator remains the same. In case of dynamic labelling, we add an extra coin-qubit to allow inter-layer flow, as shown in Fig.~\ref{fig:shift_op}.

\subsubsection{Oracle}

Finally, consider the oracle $R$ which we claim to be a controlled Grover diffusion operator (up to a phase), where control qubits are marked states. To prove this, consider the form of oracle in Eq.~\eqref{eq:search_oracle_multiple}. This operator acts trivially (as identity) on the states which does not belong to set of marked nodes $M$, while it acts as $I\otimes (I - 2|\psi_c^{(d)}\rangle \langle \psi_c^{(d)}|) $ if the state belongs to $M$. The operator $I - 2|\psi_c^{(d)}\rangle \langle \psi_c^{(d)}|= -G^{(d)}$ is Grover diffusion operator up to a phase of $e^{i\pi}$. This is results follows for the case of multilayer search oracle except the control operation is over state $|x,y,\eta_z\rangle$ where $(x,y) \in M_z$.

\begin{figure}
	\centering
	\begin{subfigure}[b]{0.45\textwidth}
		\centering
		\includegraphics[width=\textwidth]{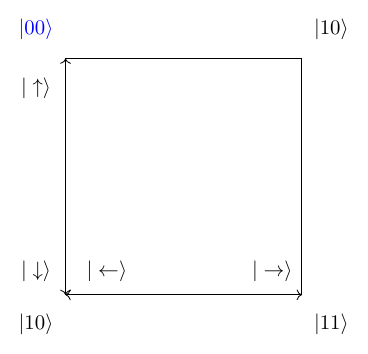}
		\caption{}
	\end{subfigure}
	\hfill
	\begin{subfigure}[b]{0.45\textwidth}
		\centering
		\includegraphics[width=\textwidth]{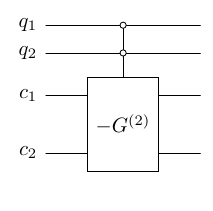}
		\caption{}
	\end{subfigure}
	\caption{(a) The qubit space for $2\times2$ lattice along with coin basis. (b) Quantum circuit implementation of Oracle for $2\times 2$ lattice. The Pauli $X$ operator is used to flip the bit to implement control operation over $|00\rangle$ position basis.}
	\label{fig:2X2_circuit}
\end{figure}

To illustrate this result, consider a quantum-walk search on $2\times 2$ lattice with a marked point chosen to be  $(0,0)$ without the loss of generality. We map the position states in qubit states as shown in Fig.~\ref{fig:2X2_circuit}. The oracle can be written as

\begin{equation}
	R = I - 2 |\psi_c^{(2)}\rangle \langle \psi_c^{(2)}|\otimes |00\rangle \langle 00 |
\end{equation}

\begin{figure}[ht]
	\centering
	\begin{subfigure}[b]{0.45\textwidth}
		\centering
		\includegraphics[width=\textwidth]{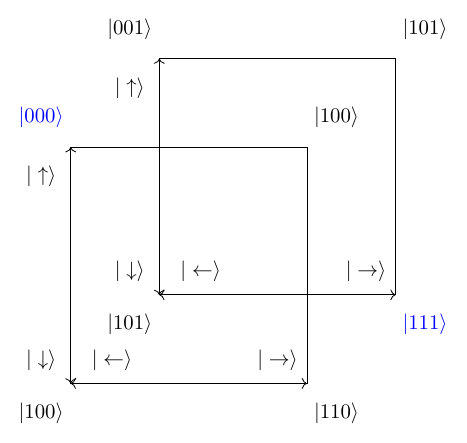}
		\caption{}
	\end{subfigure}
	\hfill
	\begin{subfigure}[b]{0.45\textwidth}
		\centering
		\includegraphics[width=\textwidth]{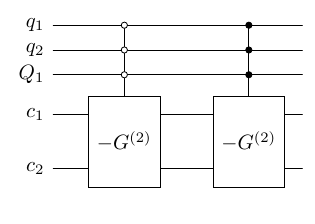}
		\caption{}
	\end{subfigure}
	\caption{(a) The qubit space for $2\times2$ lattice with two layers with coin basis. The marked nodes are shown with blue color. (b) Quantum circuit of oracle for $2\times 2$ lattice with two layers for marked nodes $|000\rangle$ and $|111\rangle$.}
	\label{fig:2X2_order_circuit}
\end{figure}

\noindent The first term doesn't affect the state and the second term only contributes when the position state is a marked state $|00\rangle$. More explicitly, 

\begin{equation}
	\begin{split}
		|0000\rangle &\rightarrow \frac{1}{2}(|00\rangle - |01\rangle - |10\rangle - |11\rangle )|00\rangle\,,\ \ \qquad \
		|0100\rangle \rightarrow \frac{1}{2}(-|00\rangle + |01\rangle - |10\rangle - |11\rangle )|00\rangle \\
		|1000\rangle &\rightarrow \frac{1}{2}(-|00\rangle - |01\rangle + |10\rangle - |11\rangle )|00\rangle\,,\qquad
		|1100\rangle \rightarrow \frac{1}{2}(|-00\rangle - |01\rangle - |10\rangle + |11\rangle )|00\rangle\,. 
	\end{split}
\end{equation}
Therefore, it's a control operation over marked state $|00\rangle$ with controlled operation $-G^{(2)}$. Figure~\ref{fig:2X2_circuit} shows the quantum circuit implementation of this Oracle. Consider the case of ordered marked points with two categories, therefore, we require one additional qubit for two layers. Further, we assume that the two categories contain marked points $|00\rangle$ and $|11\rangle$ (See Fig.~\ref{fig:2X2_order_circuit}). The oracle operator $R$ can be written as 

\begin{equation}
	R = I - 2 |\psi_c^{(2)}\rangle \langle \psi_c^{(2)}|\otimes \left[|00\rangle \langle 00|\otimes |0\rangle \langle 0| + |11\rangle \langle 11|\otimes |1\rangle \langle 1| \right]
\end{equation} 
where we assumed static labelling (but easily generalized to dynamic case). Following the similar argument as before, the oracle operator only acts non-trivially to marked labelled states which belong to set $M_z$, in this case, $|000\rangle$ and $|111\rangle$. More explicitly,

\begin{equation}
	\begin{split}
		|00\rangle|x,y,\eta_z\rangle &\rightarrow \frac{1}{2}(|00\rangle - |01\rangle - |10\rangle - |11\rangle )|x,y,\eta_z\rangle \\
		|01\rangle|x,y,\eta_z\rangle &\rightarrow \frac{1}{2}(-|00\rangle + |01\rangle - |10\rangle - |11\rangle )|x,y,\eta_z\rangle \\
		|10\rangle|x,y,\eta_z\rangle &\rightarrow \frac{1}{2}(-|00\rangle - |01\rangle + |10\rangle - |11\rangle )|x,y,\eta_z\rangle \\
		|11\rangle|x,y,\eta_z\rangle &\rightarrow \frac{1}{2}(|-00\rangle - |01\rangle - |10\rangle + |11\rangle )|x,y,\eta_z\rangle \\
	\end{split}
\end{equation}
where $|x,y,\eta_z\rangle = |000\rangle, |111\rangle$. As before, this is controlled Grover diffusion operator over marked position states acting on coin state.

\subsubsection{The complexity scaling}

In this section, we analyze the complexity scaling (resource required) of the quantum algorithm both with system size. We will focus on quantum-walk search with ordered marked points from which the complexity of the quantum tracking problem can easily be derived.

As we previously seen the Hilbert space dimensions of single and multi-layer amplification algorithms are $4mN$ and $8mN$ respectively, where $N$ is the number of lattice points and $m$ is the number of categories or layers. Therefore, the qubit requirement scale is $\mathcal{O}(\log (mN))$ with system size. The major cost of the operator in the algorithm comes from the flip-flop shift operator. We can find how many Toffoli gates required for the flip-flop shift operator for single-layer amplification 

$$4\times \sum_{n = 1}^{D/2} 16 \frac{n}{2} \approx 4(D^2 +D)  $$ 
which is $\mathcal{O}(4D^2)$. For the multi-layer amplification case, this modifies to $\approx 4(D^2 + D) + 8D'(D'+1) $ due to the additional operator needed for hopping between layers. The $n$-qubit Toffoli gate requires at least $2n$ CNOT gates \cite{shende_cnot-cost_2009}, therefore, the number of CNOT gates required is approximately of order $\mathcal{O}(8D^2)$. The construction of the oracle requires control operation of Grover's diffusion operator as many times as the number of marked points. For a large data set and a small number of marked points, we expect the cost due to shift operator to dominate, and therefore resource requirement is polynomial in the number of gates required for a single step. Therefore, the complete algorithm requires at least $\approx t_\text{op} \mathcal{O}(D^2)$ CNOT gates.

\section{Discussion and Conclusion}\label{sec:conclusion}
The conventional QWSA is aimed at finding a marked node in a graph but lacks the ability to characterize the nodes when more than one is present. We propose a modification that locates multiple marked nodes and characterizes them with respect to an existing (chronological) ordering. Clearly, our algorithm can also be extended to cases where the categorization of marked nodes is based on some attribute other than temporal. This involves the introduction of extra qubits associated with categories. We give an explicit form of oracle in two separate cases depending upon whether there's an inter-flow of probability between categories. As a concrete application, we used our formulation for particle tracking in real-time. Finally, we also construct an equivalent quantum circuit for the algorithm, with the prospect of integration with  the contemporary quantum hardware. However, this is a beginning step, where we have just scratched the tip of the iceberg, and a lot more needs to be amended in the algorithm before it gets market-ready. We will point out some immediate follow-up questions that we intend to resolve and extend the scope of the algorithm:
\begin{itemize}
	\item \textbf{More generic geometries:} We have considered the algorithm on a simple two-dimensional lattice w/o boundary conditions. The immediate generalization would be to consider more generic and perhaps non-trivial geometries with intricacies. For example, we would consider a percolation lattice in $2d$ with different weighted edges and on-site potentials. These geometries represent various scenarios in real-time systems. Another direction worth pursuing is the search on graphs themselves. A part of the problem is already addressed in Sec.~\ref{sec:quantum tracking problem} where we considered the simpler version of \cite{PhysRevLett.119.220503} where the authors discuss searches on temporal (time-varying) graphs. We would like to see if our algorithm can be extended to address multiple temporal graphs with intersecting vertices. This would be helpful in elevating the predictability of the algorithm from a tracking to a tracking-intercepting algorithm. 
	\item \textbf{Localization and Quantum State transfer:} These two concepts are seemingly disconnected. Localization explains how particle propagation (plane waves) can be restricted (localized distribution) in the presence of disorder in the media \cite{PhysRev.109.1492}. In particular, it is demonstrated in the context of DTQW in various settings \cite{zeng_discrete-time_2017,derevyanko_anderson_2018,sen_exploring_2023}. Quantum State transfer concerns the propagation of a specific quantum state from one node (origin) to another (target) through a complex network (e.g., a spin chain) \cite{PhysRevLett.78.3221,PhysRevLett.92.187902}. These two ideas are not quite connected with each other and are more disconnected from the QWSA. The question, however, is whether we can establish a connection between the search algorithm and the localization aspect by thinking of the search oracle as a disorder in an otherwise non-chaotic media. Similarly, instead of taking a complete superposition for an initial state, can we single out the marked nodes with any biased (a specific) initial state? The real question in both scenarios is to interpret the ``Search Oracle'' $R$ as a disorder operator from the physics point of view, which in turn can help understand the search oracle better and amend it for other purposes based on insights from physics.  
\end{itemize}

\section*{Acknowledgements}

H.S. thanks Kanad Sengupta for the discussion on part of the quantum circuit. K.S. was partially supported by S\~{a}o Paulo Funding Agency FAPESP Grants 2021/02304-3 and 2019/24277-8. The authors thank Prof. Urbasi Sinha for discussions and useful comments on the work. 

\section*{Author contributions statement}
K.S. conceived the research problem. K.S. and H.S. jointly formulated the quantum algorithms and conducted independent numerical experiments. H.S. proposed the quantum circuit with guidance and support from K.S. H.S. took the lead in manuscript preparation with assistance from K.S.

\section*{Data availability}
The datasets generated and analysed during the current study are publicly available via \href{https://github.com/Kmax-art/q-search}{q-search repository}. The supplementary material contains a .gif movie for tracking problem. 

\section*{Additional information}
\textbf{Competing financial interests:} The authors declare no competing financial interests.

\vfill

\newpage

\bibliographystyle{quantum}
\bibliography{sample}

%
%
%
%

\end{document}